# Polarization-Aligned, Spectrally Consistent Quantum Emitters in As-Exfoliated Carbon-Doped Hexagonal Boron Nitride

*Sofiya Karankova,[1,2] Yeunjeong Lee,[1,3] Seungmin Park,[4,5] Kenji Watanabe,[6] Takashi Taniguchi,[7] Jin-Dong Song,[1,8] Young Duck Kim, [\*,4,5] Yong-Won Song, [\*,1,8] and Hyowon Moon[\*,1,8]*

S. Karankova,[1] Yeunjeong Lee,[1] Jin-Dong Song,[1] Yong-Won Song,[1] Hyowon Moon[1]

[1]Center for Quantum Technology, Korea Institute of Science and Technology, Seoul 02792, Republic of Korea

S. Karankova[2]

[2]Division of Nano & Information Technology, KIST School, University of Science and Technology, Seoul 02792, Republic of Korea

Y. Lee[3]

[3]Department of Physics, Korea University, Seoul 02841, Korea

Seungmin Park,[4,5] Young Duck Kim,[4,5]

[4]Department of Physics, Kyung Hee University, Seoul 02447, Republic of Korea

[5]Department of Information Display, Kyung Hee University, Seoul 02447, Republic of Korea

K. Watanabe[6]

[6]Research Center for Electronic and Optical Materials, National Institute for Materials Science, Tsukuba 305-0044, Japan

T. Taniguchi[7]

[7]Research Center for Materials Nanoarchitectonics, National Institute for Materials Science, Tsukuba 305-0044, Japan

Jin-Dong Song,[8] Yong-Won Song,[8] Hyowon Moon[8]

[8]Nanoscience and Technology, KIST School, University of Science and Technology, 02792 Seoul, Republic of Korea

E-mail: hwmoon@kist.re.kr, ysong@kist.re.kr, ydk@khu.ac.kr

Solid-state quantum emitters constitute an essential building blocks of integrated quantum photonic circuits. Among potential emitter platforms, hexagonal boron nitride (hBN) hosts single-photon emitters in an atomically thin lattice amenable to photonic integration. However, multi-step fabrication approaches, limited defect specificity, and poor emission wavelength repeatability limit the performance of hBN quantum light sources relative to established solid-state architectures. Developing methods to induce emitters that are both suitable for planar photonic devices and that exhibit consistent optical properties remains a key objective. In this work, we identify quantum emitters in as-exfoliated carbon-doped hBN that exhibit both stable and repeatable emission energies together with polarization-aligned dipoles. Owing to the high lattice crystallinity, these single-photon light sources demonstrate exceptional spectral stability with a standard deviation of 7 μeV. The emission energy is reproducible and confined within a narrow range of 2.2825 ± 0.0042 eV. Notably, consistent dipole alignment for absorption and emission polarization suggests that the intrinsic defects are of the same nature. The color centers are observed in as-exfoliated hBN without any post-treatment, significantly facilitating further interfacing with planar photonic structures. These reproducible, polarization-aligned quantum emitters in as-exfoliated hBN provide a versatile platform for scalable integration, offering a pathway toward a broad range of quantum technologies.

# 1. Introduction

Solid-state quantum emitters (QEs) provide a practical route toward scalable quantum photonic technologies that can operate under ambient conditions and be integrated into compact device platforms[1,2]. In particular, two-dimensional (2D) materials offer unique characteristics as atomically-thin layered crystals that can be transferred directly onto diverse photonic structures and substrates, facilitating efficient coupling to optical modes and simplifying integration in quantum photonic circuits. Among these materials, hexagonal boron nitride (hBN) has attracted significant attention due to its wide bandgap, which can host color centers with emission energies ranging from ultraviolet (UV) to infrared and is independent of layer number or crystal geometry[3]. Notably, point defects in hBN can act as single-photon sources and have been reported to be among the brightest quantum light sources to date[4], exhibiting narrow and tunable zero-phonon lines (ZPLs) at room temperature (RT)[5,6] which are essential for efficient cavity coupling and photon indistinguishability[7-10]. More recently, point defects in hBN have been shown to host optically addressable spin states, enabling applications in quantum sensing including nanoscale magnetic, temperature, strain, and pressure sensing[11-17].

Various fabrication techniques have been proposed to create QEs in hBN, including nanoindentation[18,19], nanostructure patterning[20], irradiation with an electron-beam[21-23], pulsed laser[24], and focused ion beam[5,25]. However, these approaches inevitably lead to local lattice damage which can induce stronger background emission that degrades single-photon purity. They also often require additional processing steps such as high-temperature annealing to activate or stabilize the emitters, thereby increasing fabrication complexity. Other treatments applied to exfoliated hBN flakes such as thermal annealing[26], plasma exposure[27], chemical etching[28], and ion implantation[29] often result in limited emitter homogeneity due to unwanted chemical changes and lattice disorder. In contrast, hBN grown by chemical vapor deposition can host high densities of QEs with a minimum emission wavelength spread of ~20 nm[30-32], though the host material quality remains limited, as evidenced by Raman spectroscopy[33]. Overall, despite significant progress, many existing fabrication approaches struggle to simultaneously satisfy all key requirements necessary for reliable operation and further integration. Addressing this challenge requires approaches that combine simplicity, compatibility with diverse surfaces, and the reproducible generation of single-photon emitters with uniform and high-quality optical characteristics.

In this article, we report narrowband single-photon sources emitting at 2.28 eV from as-exfoliated carbon-doped (C-doped) hBN exhibiting excellent temporal and spectral stability and collective dipole polarization alignment. Notably, the emitters are present in as-exfoliated flakes without the need for any treatments, simplifying fabrication and minimizing material and photophysical degradation. Therefore, these emitters show low background emission levels owing to the high crystalline quality of the hosting material and reduced noise sources from the local environment. By systematically analyzing the photophysical properties of the emitters, this work establishes a robust experimental foundation for reproducible quantum light sources readily compatible with integrated photonic architectures.

## 2. Results and Discussion

### 2.1. Carbon-doped hBN by HPHT

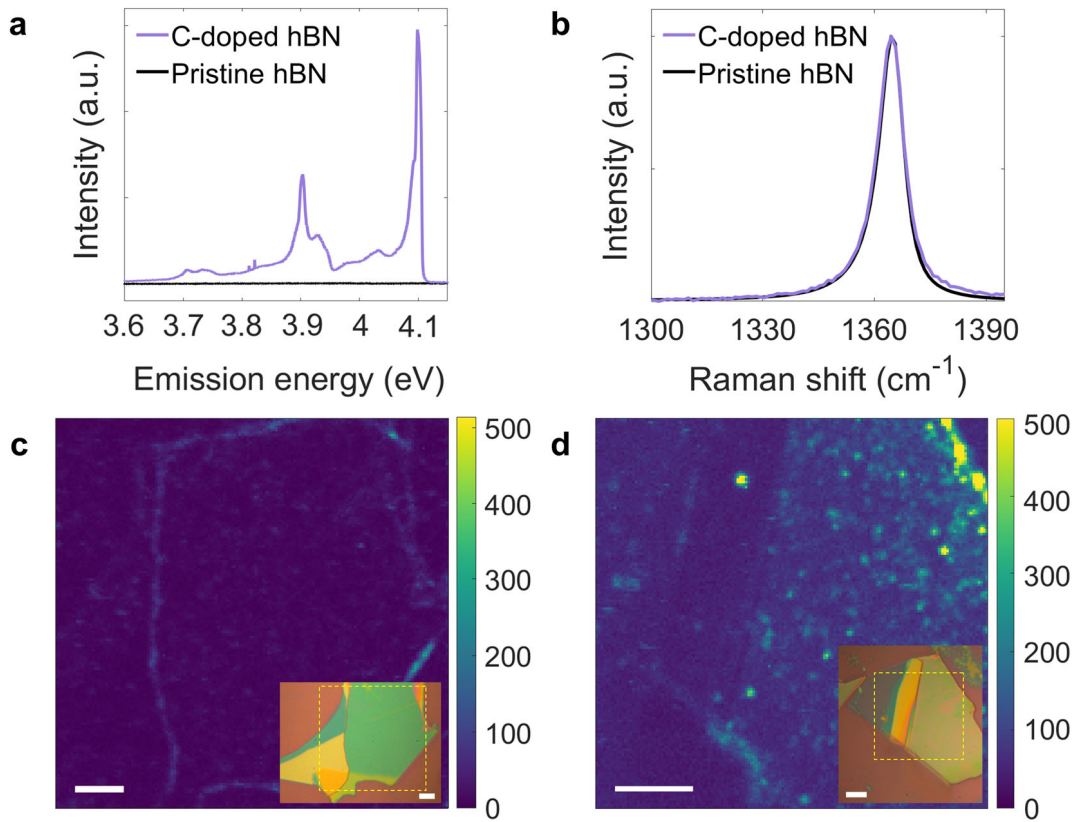

**Figure 1.** Optical characterization of hBN. a) UV PL from pristine and C-doped hBN crystals recorded at 5 K, where the ZPL peak of the UV emitter ensemble is at 4.0989 eV. Strongly coupled LO phonon modes with sharp peaks are red-shifted by 195.7 meV and 392.5 meV from the ZPL, respectively. b) Raman characterization of pristine and C-doped hBN demonstrates the high crystallinity of hBN flakes synthesized via the HPHT method. The position of the $E_{2g}$ phonon mode is observed at 1364.4 cm$^{-1}$. FWHM linewidths of the pristine and C-doped hBN are 8.2 and 8.5 cm$^{-1}$, respectively. c) and d) Exemplary PL maps of pristine and C-doped HPHT hBN flakes exfoliated on top of a Si/SiO$_2$ substrate, respectively (insets: optical microscope images of the hBN flakes). Scale bar: 5 μm.

We exfoliate hBN flakes from bulk crystals of pristine and C-doped hBN on top of Si/SiO$_2$ substrates. C-doped hBN crystals were fabricated by high-pressure high-temperature (HPHT) synthesis followed by annealing with a graphite susceptor after the synthesis (see details in Methods). **Figure 1a** shows the distinctive spectral features in the UV photoluminescence (PL) spectrum of C-doped hBN in contrast to pristine hBN[34-36]. Sharp ZPL peaks of the UV emitter ensemble at 4.0989 eV and strongly coupled longitudinal optical (LO) phonon modes with the red-shifts of 195.7 meV and 392.5 meV confirm the existence of carbon defects in the flakes[34,36,37]. We mainly studied thick flakes with thickness above 100 nm (**Figure S1**), which are readily available for integration with photonic structures[38-40]. The $E_{2g}$ phonon mode of pristine and C-doped hBN from Raman spectroscopy reveals high crystal quality of both materials (**Figure 1b**, details in Methods). The mode appears at 1364.4 cm$^{-1}$ with the full width at half maximum (FWHM) linewidths of 8.22 and 8.48 cm$^{-1}$ for pristine and C-doped hBN, both showing the lower end of reported values and indicating intact crystalline quality with only minor lattice perturbation[32,33,41].

**Figure 1c** and **1d** show the PL signal in the visible spectral range from the pristine and C-doped hBN flakes by using a home-built confocal setup (see Methods), while the insets show the OM photographs of the corresponding flakes. The pristine hBN exhibits no detectable emission centers in the visible spectral range, confirming the absence of optically active defects. In contrast, the C-doped hBN reveals a high density of bright, spatially localized emission spots. These color centers are readily distinguishable due to their strong intensity contrast against the background. Within the C-doped sample, the emission intensity varies from emitter to emitter, which can be attributed to differences in defect depth. Individual quantum emitters are mostly isolated and selectively addressable with sufficient density, which is particularly advantageous for subsequent integration with photonic or plasmonic platforms. In the following section, we discuss the RT photophysical properties of a representative group of single-photon emitters identified in the C-doped HPHT hBN flakes.

## 2.2. Room-temperature photophysics of QEs in C-doped hBN

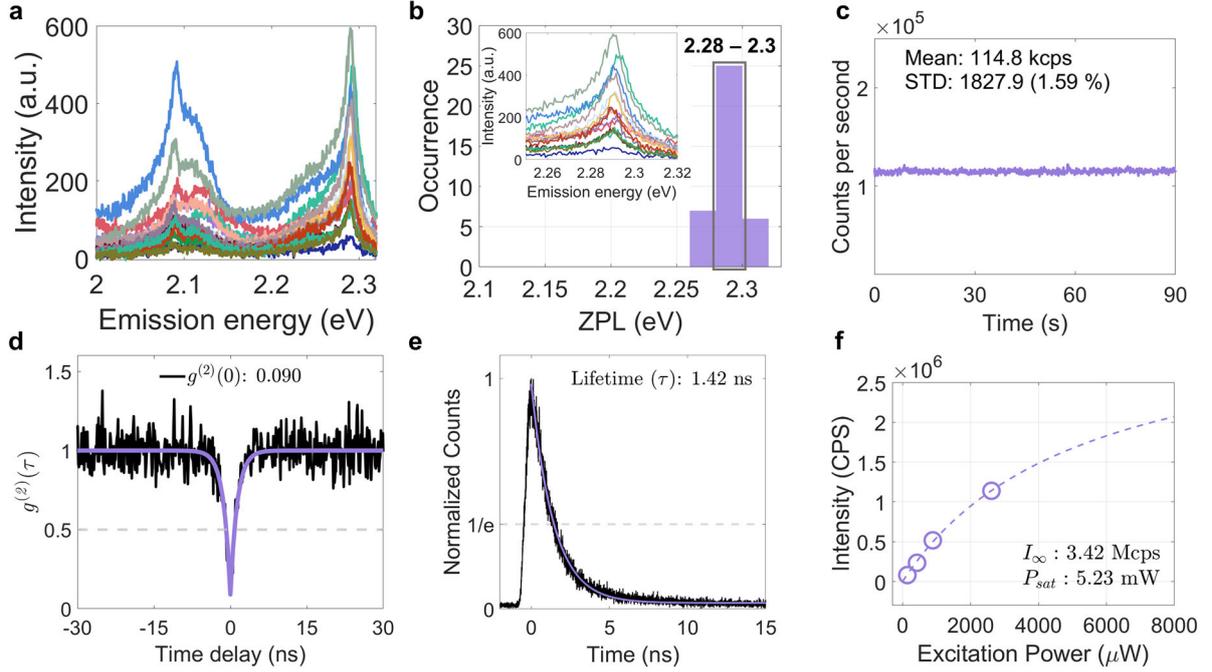

**Figure 2.** Room-temperature characterization of QEs in as-exfoliated C-doped hBN. a) PL spectra recorded over 16 QEs consisting of a narrow repeatable ZPL and a PSB with a distinct LO phonon mode. b) ZPL distribution histogram over 38 emitters from various hBN flakes (inset: narrowband PL spectra of ZPLs from Figure 2a, bin size: 0.02). c) Time-dependent PL intensity showing a continuous, stable operation of a QE at 114.8 ± 1.8 kcps. d) Second-order correlation function with an extracted photon purity of $g^{(2)}(0) = 0.09$ recorded within a 25-nm wavelength range. e) Time-resolved PL decay measurement yielding the QE lifetime of 1.42 ns. f) Power-dependent measurement yielding $I_\infty$ and $P_{sat}$ of 3.42 Mcps and 5.23 mW, respectively.

Repeatability of the emission energy is a distinct feature of the discovered group of QEs from untreated C-doped hBN. We plot the exemplary PL spectra of 16 intrinsic defects in **Figure 2a**. A representative spectrum features a narrow ZPL peaking at 2.2846 eV, accompanied by a phonon sideband (PSB) centered at 2.0913 eV with a distinct LO phonon mode at 2.083 eV. Consequently, we collect statistics from a total of 38 QEs from various as-exfoliated C-doped hBN flakes and plot a histogram with the ZPL energies in **Figure 2b**, where the magnified spectra at the ZPL energy of the emitters in Figure 2a are shown as an inset. The consistent spectral features compared to conventional visible hBN QEs, which exhibit a very narrow ZPL emission energy distribution at 2.290 ± 0.014 eV across multiple flakes, highlight the reproducibility of the defects and their local environment. All recorded data from the investigated areas were included, with the only filtering criterion being the presence of $g^{(2)}(0) < 0.5$ to ensure the quantum nature of the color centers. In addition to the ZPL consistency, we observe identical phonon features within the PSB, including a sharp line at 2.083 eV (Figure 2a). This peak is attributed to the dominant LO phonon mode, red-shifted by 202 meV from the ZPL[42,43]. Such high energy repeatability indicates that the emitters likely originate from the same family of C-induced defects. To support this theory, we probe another C-doped hBN flake and find a large density of QEs with matching LO mode energies to the initial group of QEs that we discuss in our work (see **Figure S2**).

We characterize the emitter through time-resolved PL and photon correlation measurements to assess its emission stability, single-photon purity, and excited-state lifetime. A time-dependent PL intensity in **Figure 2c** reveals minimal variations of emission of 114.8 ± 1.8 kilocounts per second (kcps) under a laser excitation power of 100 μW, indicating stable operation of a QE without bleaching or blinking behavior[21]. We also provide longer stability measurements for 5 minutes in **Figure S3** to emphasize the exceptional stability of the QE with negligible energy fluctuations highlighted by a standard deviation (STD) of 151 μeV. The representative second-order autocorrelation function is shown in **Figure 2d**, demonstrating high purity of the quantum light source with a $g^{(2)}(0)$ of 0.09 recorded within a 25-nm wavelength range. We measure the $g^{(2)}(0)$ values from other emitters within the 85-nm filtered spectral range and plot the histogram (see **Figure S4**). All investigated defects exhibit $g^{(2)}(0) < 0.5$ without background correction, with more than half (57%) of the emitters showing high purity with $g^{(2)}(0) < 0.2$ even at RT, confirming low background signal due to the high crystallinity of the HPHT-synthesized host crystal. Time-resolved PL decay yields a typical lifetime of 1.42 ns (**Figure 2e**), in excellent agreement with previous reports[7]. A histogram of the lifetime distribution for each corresponding QE, derived from either the $g^{(2)}(\tau)$ or the fitted PL decay, can be found in **Figure S4**. Together with the high-purity single-photon emission, these results indicate a well-isolated, stable two-level system. Visible quantum emitters in hBN are also well-known for their exceptional brightness[4,5]. A power saturation measurement in **Figure 2f** shows the saturation intensity $I_\infty$ and saturation power $P_{sat}$ of 3.42 Mcps and 5.23 mW, respectively, by fitting to a saturated-emitter model $I = I_\infty \times P/(P + P_{sat})$[44].

    The results demonstrate that QEs in C-doped hBN exhibit consistent emission energies, indicating a high degree of uniformity in their optical properties. We attribute it to the highly ordered crystalline environment of as-exfoliated hBN with minimal strain and chemical contamination. Additionally, the emission from C-doped hBN defects features a narrow ZPL with a FWHM of 11.6 meV (2.7 nm), among the lowest values reported for RT hBN QEs[45,46]. These spectral characteristics, together with high single-photon purity, stable operation without blinking, and bright emission due to nanosecond-scale excited-state lifetimes, make the quantum emitters exhibit highly reliable photophysical characteristics in a two-dimensional host material.

## 2.3. Low-temperature measurements of QEs in C-doped hBN

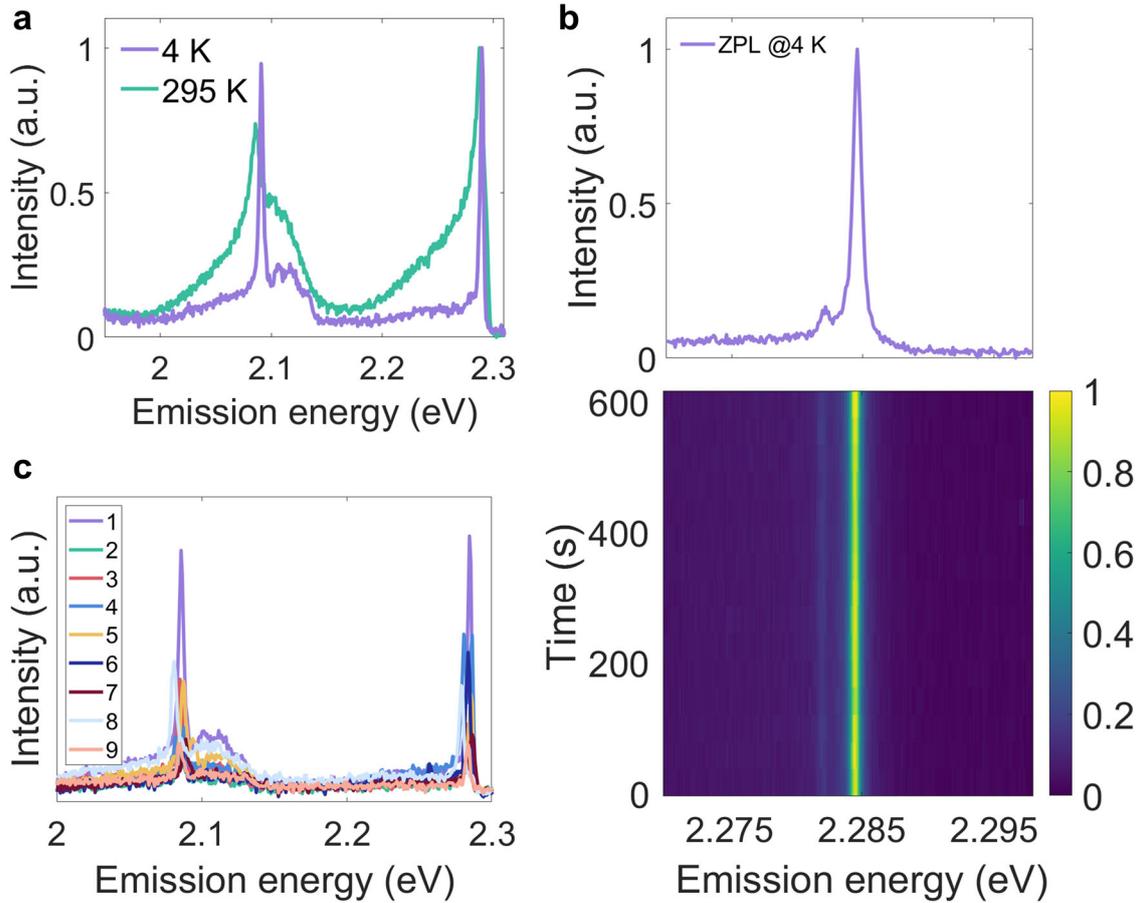

**Figure 3.** Low-temperature characterization of QEs in as-exfoliated C-doped hBN. a) PL spectra of a representative emitter at 295 K and 4 K with subtracted background and normalized for comparison purposes. b) PL spectra of the ZPL centered at 2.2845 eV with a FWHM of 0.8 meV at a higher 1200 lines/mm spectrometer grating. Bottom graph – corresponding time-resolved PL spectra of the peak recorded for 10 minutes. STD was determined as 7 µeV, indicating minimal spectral diffusion of the ZPL. c) Exemplary PL spectra of nine QEs. Mean and STD values were extracted to be 2.2825 ± 0.0042 eV, respectively.

Cryogenic measurements of QEs in C-doped hBN at 4 K reveal a narrow and extremely stable ZPL with negligible spectral diffusion, underscoring their potential suitability for practical quantum photonics applications. **Figure 3a** shows the wide-range PL from the same defect at RT and low-temperature (LT) (see setup arrangement in **Figure S5**). A temperature-induced 0.4 meV blue shift of the spectrum at 4 K relative to 295 K is attributed to reduced anharmonic phonon-phonon and electron-phonon interactions at LT[47]. Significant broadening of the total spectrum at RT (295 K) presumably originates from both fast dephasing (homogeneous broadening) and local electrostatic environment (inhomogeneous broadening) due to the close proximity to the surface. Such effects can be significantly suppressed at cryogenic temperatures as can be seen from a narrow-band ZPL at LT (4 K). The spectrum comprises a ZPL peak at 2.2845 eV and a sharp LO phonon peak at 2.0861 eV of similar intensity. Strong coupling to LO phonons is present in both RT and LT measurements, which is typical for the carbon-related defects[37,42,48].

A common issue that many hBN-hosted QEs face is emission line instability that worsens over time. The stability of a quantum light source is essential for practical quantum applications; however, it remains particularly challenging to realize it for QEs in two-dimensional materials. The main reason lies in the low thickness of the material, which allows nearby electrostatic charges to influence the QE, and in defect migration within the crystal lattice upon laser excitation[40]. Together, these factors give rise to such effects as spectral diffusion and luminescence intermittency[40]. Although approaches including annealing and encapsulation can provide a noticeable reduction in spectral diffusion[19,49,50], the resulting emission stability remains insufficient for the practical use of these QEs in quantum technologies. Given the importance of this issue, we quantitatively evaluate the spectral diffusion of our C-doped hBN QEs. The PL spectrum of the ZPL as a function of time using a higher 1200 lines/mm spectrometer grating is demonstrated in **Figure 3b** (data from the LO phonon line in **Figure S6**). A ZPL FWHM of 0.8 meV is obtained under non-resonant excitation and is close to the spectrometer resolution (**Figure S7**). The intrinsic linewidth is expected to be narrower and could be resolved under resonant excitation[10,51]. Spectral diffusion of the peak is 7 μeV, which is the lowest value among the QEs in hBN reported so far, significantly below the current state-of-the-art value of 45 μeV from the B-center at a much shorter wavelength[21,31,52].

The ZPL energy is also highly consistent at LT across the range of emitters (**Figure 3c**). We confirm the quantum nature of each studied defect with a second-order autocorrelation measurement (**Figure S8**). Very low inhomogeneity with an STD of 4.2 meV is another unique feature of our emitters, being on a similar scale to previous state-of-the-art studies on large-area QE fabrication in hBN[30,31,45,52,53]. Such closely spaced emission energies provide a favorable starting point for fine spectral tuning, thereby facilitating spectral matching required for multi-emitter quantum interference experiments. The achieved results provide strong evidence that the emitters originate from a similar C-related defect type, which is also supported by the reproducibility of ZPL–LO phonon energy difference (**Figure S9**) and by the polarization measurements described in the next section.

## 2.4. Polarization measurements of QEs in C-doped hBN

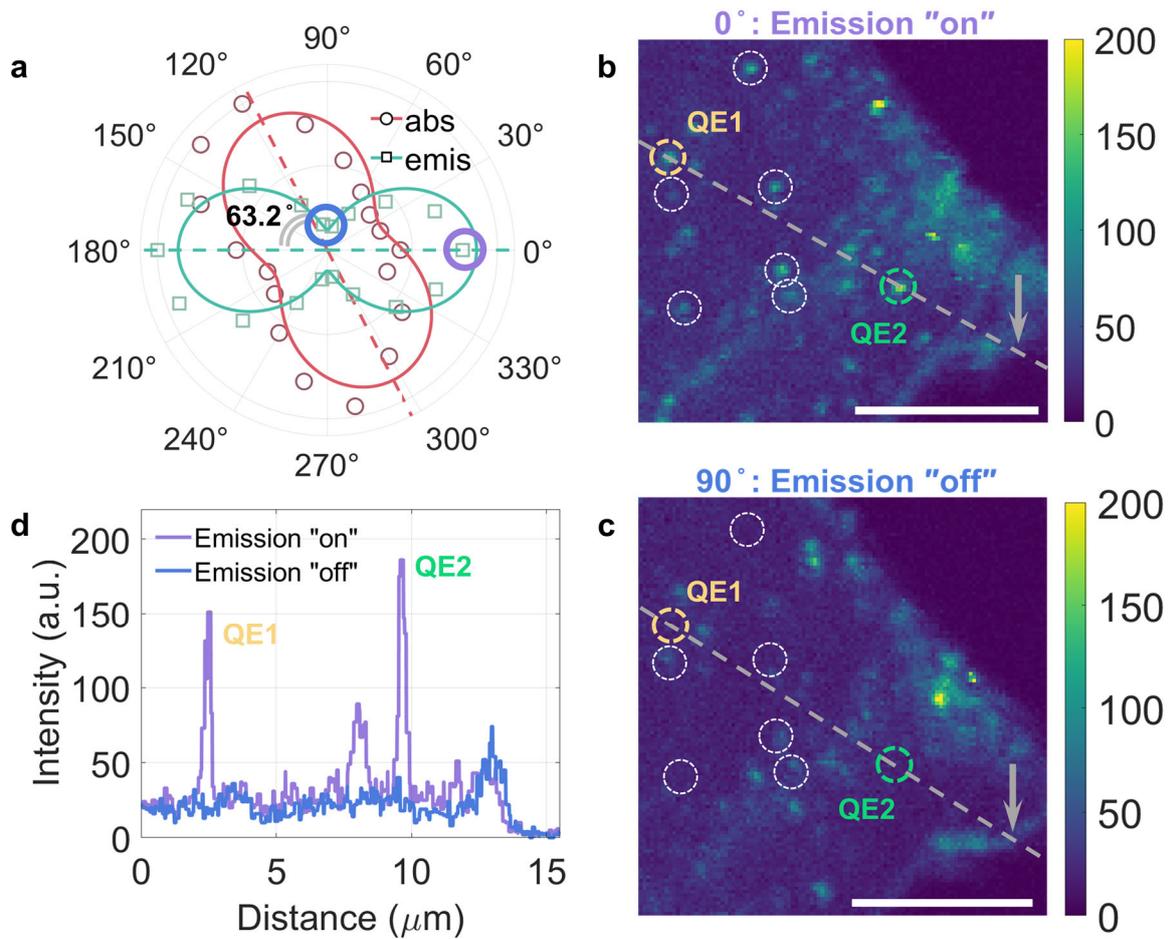

**Figure 4.** Group polarization measurements of QEs from C-doped hBN. a) Absorption (abs, red) and emission (emis, teal) dipole polarization of the exemplary defect, labeled as QE2 in the following figures show dipole-like angular dependence. Maximum (0˚, reference axis) and minimum (90˚) intensity points for emission polarization are marked with purple- and blue-colored circles, respectively. b) and c) A set of PL maps corresponding to maximum and minimum emission polarization, scale bar 5 μm. Positions of eight QEs are indicated in white circles; among them, two representative QEs are highlighted with colored circles. For reference purposes, a gray arrow marks the edge of the hBN flake. Emitters demonstrate alignment in polarization and simultaneously switch from "on" to "off" states. d) Line cuts along the gray dashed lines from the PL maps on Figure 4b,c demonstrate the modulation of the emitters intensity, whereas the background intensity and hBN flake remain constant over the entire range of polarization angles.

Control of the emitter polarization is essential for polarization-matched multi-photon interference. **Figure 4a** shows the polarization properties of the individual emitter labeled as QE2 by rotating a half-wave plate in the excitation and detection paths, respectively. The emission maximum (in purple circle) is defined as 0°, and this reference axis is consistently used throughout the subsequent polarization-resolved measurements. The polar plot clearly exhibits dipole-like angular dependence for both absorption and emission, where the dipole axes differ by 63.2°. The involvement of distinct transition dipole moments indicates participation of multiple electronic states, consistent with previous reports in hBN[7,54,55]. In addition, the angle difference close to 60° suggests a possible relationship between the dipole orientation and the crystallographic lattice directions of hBN[52].

Beyond single-emitter behavior, polarization-resolved PL maps reveal a striking collective response across spatially separated emitters. When the detection polarization is rotated from 0° to 90°, multiple quantum emitters simultaneously switch between high- and low-intensity states, as shown in the dashed circles in **Figure 4b** and **4c**. The intensity profile extracted along the dashed line in **Figure 4d** confirms concurrent suppression of QE1 and QE2 at 90°, while the flake-edge emission and background noise signal remain unchanged, excluding global attenuation effects. **Figure S10** presents a similar observation for absorption polarization. **Figure S11** shows the polarization orientation of the emitter labeled as QE1, proving that this emitter also shares the same characteristics. This synchronized modulation demonstrates a common emission polarization axis among the emitters, potentially related to defect dipoles aligned with the lattice. Moreover, such collective response provides strong evidence that the emitters are of the same nature, suggesting an identical defect structure within the group. Accordingly, we propose carbon-oxygen defect complexes as a possible microscopic origin of the emitters discussed in this work[52,56,57].

Importantly, these observations are obtained from as-exfoliated flakes without any post-processing treatments. The absence of destructive post-processing such as irradiation or annealing suggests minimal lattice damage, allowing the intrinsic crystal symmetry to govern defect dipole orientation. We therefore attribute the observed polarization alignment and reproducibility to a well-preserved hBN lattice. Such collective alignment of dipole orientations is particularly significant, as polarization matching is a prerequisite for multi-photon interference experiments, including Hong–Ou–Mandel interference.

## 3. Conclusion

We demonstrate that intrinsic color centers in as-exfoliated C-doped HPHT hBN give rise to single-photon emitters with reproducible and narrowly distributed ZPL energies centered at 2.28 eV, together with stable quantum emission in both spectral and temporal domains. The observed consistency in emission energy and dipole polarization across multiple emitters indicates a common defect origin, offering a pathway toward resolving a long-standing challenge of defect specificity and emitter-to-emitter variability in hBN-hosted quantum emitters and representing an important step toward indistinguishable photon generation. Importantly, these QEs emerge without any post-exfoliation activation or processing, directly linking their optical uniformity to the intrinsic high crystalline quality. Our finding eliminates key fabrication bottlenecks associated with multi-step treatments and stochastic defect activation, offering a practical and materials-level solution for reproducible quantum light sources. Together, our results establish C-doped hBN as a distinct and robust emitter platform, advancing hBN quantum emitters from isolated demonstrations toward scalable and reliable components for integrated quantum photonic systems.

## 4. Methods

*Sample preparation*

C-doped hBN was made by annealing pristine hBN at 2000 ℃ for 1 hour in a high-frequency furnace with a graphite susceptor. Both pristine and C-doped hBN were later exfoliated with a tape onto patterned $SiO_2$/Si substrates of size 10 × 10 mm. No additional treatment was conducted.

*Room-temperature measurements*

Raman spectra were recorded under a few mW power with 10 accumulations with 5 sec acquisition per accumulation (Renishaw, inVia).

A lab-built PL confocal setup was used to acquire PL maps, spectra, lifetime, and anti-bunching measurements at room temperature. Multiple CW and pulsed laser sources with wavelengths of 450, 470, 514, and 532 nm were employed depending on the experiment specifics, with the excitation beam focused onto the sample using an objective lens (100×, NA = 0.9, MPLFLN100X, Olympus). The system incorporated a 2D galvo scanner (GVS002, Thorlabs) in a 4f configuration, and OM imaging was enabled by coupling a white-light source into the excitation path and imaging with a CMOS camera. Polarization-resolved measurements were performed using polarizers in both excitation and collection arms together with a half-wave plate. Collected emission was directed through a 50:50 beamsplitter and a long-pass filter to cut the laser before passing through a pinhole placed at the conjugate image plane. A flipping mirror routed the signal either to a spectrometer (MonoRa512i, DongWoo Optron) equipped with a CCD detector (iDUS 401, DU401A-BVF, Andor) or to a free-space Hanbury Brown-Twiss interferometer with two avalanche photodiodes (SPCM-AQRH-44, Excelitas) for photon-correlation measurements. The ZPL energy was extracted by fitting the PL peak with a Lorentzian function, and laser fitting was performed with a Voigt function fit. Second-order autocorrelation histograms with 128 ps resolution were recorded using a time-correlated single-photon counting module (Picoharp300, PicoQuant). The quantum efficiency of the setup cannot be evaluated reliably and is therefore beyond the scope of this work. Data from Figure 2 were recorded with a 532 nm laser (MGL-III-532, CNI) unless otherwise stated. The PL map (120*120 pixels, dwell time – 5 ms/pixel) from Figure 2a was acquired in the wavelength range 540 – 625 nm at an excitation power of 285 µW. The acquisition time of the PL spectrum in Figure 2b was set to 10 s. In Figure 2c, the ZPL wavelengths were acquired with various lasers including 470 nm, 514 nm, and 532 nm. Positions of QEs did not overlap, and each PL spectrum corresponds to a separate QE. For each PL spectrum we confirmed $g^2(0) < 0.5$ to ensure the quantum nature of the defect. Brightness was recorded using the 550-nm long-pass filter. The anti-bunching curve was measured at a power of a 1 mW. The temporal stability trace was recorded with a resolution of 100 ms at 100 µW power. Lifetime measurements were recorded using a 470 nm pulsed laser (laser driver is PDL 800-D model, laser head – D-C-470 model) with a power of 180 µW and 4 ps resolution using the Picoharp300. All equipment for lifetime measurements was from PicoQuant.

*Low-temperature measurements*

A confocal setup with an identical configuration to that for the RT measurements was built using a closed-cycle cryostat (Cryostation s100-CO, Montana Instruments). The target temperature for all the measurements was fixed at 4 K. A continuous-wave 532 nm laser (MGL-III-532, CNI) served as the excitation source. The beam was focused onto the sample using a 100× objective (NA = 0.9, EC Epiplan-NEOFLUAR 100×, ZEISS) and scanned with a two-dimensional galvo scanner (GVS002, Thorlabs) integrated into a 4f relay-lens system. The excitation power for each laser was measured before the objective with a calibrated power meter (PM100USB, Thorlabs). Optical images were acquired within the same confocal setup with a white-light source (MNWHL4, Thorlabs) coupled to the excitation path and capturing the reflected signal using a CMOS camera (CS165CU1/M, Thorlabs). The emitted

PL was collected through the same objective and directed to the detection arm via a 50:50 non-polarizing beamsplitter. We used a 540-nm bandpass filter to block laser light, and a 650 nm shortpass filter to limit the background luminescence from longer wavelengths. A flipping mirror enabled switching between two detection routes: (i) a spectrometer (MonoRa750i, DongWoo Optron) with a CCD detector (iDus 401, DU401A-BVF, Andor) for acquiring spectra, and (ii) a free-space Hanbury Brown–Twiss interferometer composed of a 50:50 beamsplitter and two avalanche photodiodes (SPCM-AQRH-44, Excelitas) for second-order autocorrelation measurements. Photon-correlation traces were recorded using a time-correlated single-photon counting module (PicoHarp 330, PicoQuant).

Data from Figure 2 were recorded with a 532 nm laser (MGL-III-532, CNI) unless otherwise stated. The antibunching curve from Figure 4b was measured in the range 540-625 nm at 285 µW power at RT after matching the position of the QE with LT measurements. For Figure 4d, a sampling frequency of 0.05 Hz with 20 s acquisition time for 30 acquisitions was used for the phonon line, and a sampling frequency of 0.025 Hz with 40 s acquisiyion time for 15 acquisitions was used for the ZPL. Lorentzian fitting was used for each line cut. Data from Figure 1b were acquired with a 193-nm laser at a power of 22 µW. The PL spectrum was recorded with a 10 s acquisition time.

**Acknowledgement**

This work was supported in part by a National Research Foundation of Korea (NRF) grant funded by the Korea government (MSIT) (No. RS-2022-NR068228, No. RS-2023-00210097, and No. RS-2025-02310566), in part by National Research Council of Science & Technology (NST) grant (GTL25011-000) funded by the Korea government (MSIT), and in part by the Institutional Program (No. 26E0002) funded by the Korea Institute of Science and Technology (KIST), Republic of Korea.

K.W. and T.T. acknowledge support from the JSPS KAKENHI (Grant No. 21H05233 and No. 23H02052), the CREST (JPMJCR24A5), JST, and World Premier International Research Center Initiative (WPI), MEXT, Japan.

**Author Contributions**

S.K., Y. D. K, Y.-W. S, and H. M. concieved the experiment and wrote the paper with the input from all authors. S. K. performed the photophysical measurements, analysed the data, and prepared the manuscript. Y. L. and J.-D. S. helped to prepare the sample and build the measurement system. S. P. and Y. D. K. measured and analysed UV chareacteristics of hBN. K.W. and T.T. prepared pristine and c-doped hBN HPHT crystals.

**Competing Interests Statement**

The authors declare no competing interests.